# Wave analysis in the complex Fourier transform domain:
# A new method to obtain the Green's functions of dispersive linear partial differential equations

Minjiang Zhu[*]

*Department of Mechanics and Aerospace Engineering, Southern University of Science and Technology, Shenzhen, Guangdong, 518000, China*

**Abstract**

**This paper provides a new analytical method to obtain Green's functions of linear dispersive partial differential equations. The Euler-Bernoulli beam equation and the one-dimensional heat conduction equation (dissipation equation) under impulses in space and time are solved as examples. The complex infinite-domain Green's function of the Euler-Bernoulli beam is derived. A new approach is proposed to obtain the finite-domain Green's function from the infinite-domain Green's function by the reflection and transmission analysis in the Complex Fourier transform domain. It is found that the solution obtained by this approach converges much better at short response times compared with the traditional modal analysis. Besides, by applying the geometric summation formula for matrix series, a new modal expansion solution requiring no calculation of each mode's inner product is derived, which analytically proves the wave-mode duality and simplifies the calculation. The semi-infinite-domain cases and the coupled-domain cases are also derived by the newly developed method to show its validity and simplicity. It is found that the 'non-propagating waves' also possess wave speed, and heat conduction can also be treated as propagating waves.**

**Keywords:**

**Time-dependent Green's function, wave-mode duality, complex Fourier transform domain,**

---

[*] Corresponding author: Email address: SUSTech: 11713022@mail.sustech.edu.cn    UIUC: mz51@illinois.edu

**self-similar function, semi-infinite beam, coupled beam.**

# 1 Introduction

A vibrating system can be studied in terms of either vibration modes or propagating waves [1]. The coexistence of these two types of solutions is called the wave-mode duality. In the mode approach, free vibration is mathematically treated as an eigenvalue problem, and forced vibration is analyzed as a linear combination of the eigenfunctions (vibration modes). The mode method can be found in any structural dynamics textbook, thus not further introduced. The vibration of elastic structures can also be described as waves propagating and decaying in waveguides. For dispersive wave equations, the dispersive relation between frequency and wavenumber is not proportional, which means waves under different frequencies travel at different speeds. Waves under a single frequency are called monochromatic waves. Monochromatic wave reflection and transmission can be simply described in the spatio-temporal domain when incident upon discontinuities. Mace and Mei investigated monochromatic wave reflection and transmission in Euler-Bernoulli [2] and Timoshenko beams [3]. Fahy, Gardonio [4], and Wang et al. [5] superimposed monochromatic waves on beams considering infinite reflections to prove the wave-mode duality. If the applied load contains discrete frequency information, analyses for each frequency component are applied, and responses under each frequency are superposed to obtain the final response.

However, wave analysis becomes tricky for dispersive equations when the applied loadings are not periodic. The time-dependent Green's function (GF) is sought in such cases. The physical interpretation of GF is the response of a linear system under an impulse load; GF is also so important mathematically because the response of such a system under an arbitrarily applied load can be obtained by integral transformation with its GF. The GF can be simply obtained by modal expansion. Such solutions converge poorly at short response times, as high-frequency terms are truncated. Such error is compensated by low-frequency terms at relatively long times, yet at short times when low-frequency waves have not reached the point of interest, such errors are exposed. When extreme short-time responses are required, boundary effects can be neglected as most waves have not reached the boundaries, let alone reflected, and the beam can be treated with an infinite scale. Graff [1] derived the exact self-similar infinite-beam GF. This analytical expression is widely used for the short-time-response comparison (e.g., Buessow [11]). However, this single-term function diverges fast as time increases. Semi-infinite-domain solutions are preferred. Basile and Sébastien [6] discovered that thin, brittle rods (such as spaghetti) seldom break in half because of the self-similar wave reflection, and they well predicted the time and the location of breaks on the clamped-free rods by the method of images, an approximation of wave reflection. Akkaya and Horssen [7] investigated the exact GF of semi-infinite Euler-Bernoulli beams by finding the reflected terms satisfying specific boundary conditions, especially for damping boundaries. Their work partially compensates for the inaccuracy of the infinite-beam GF, yet in the real world, a semi-infinite beam is just an extreme case when the impulse is close to one boundary and far from the other; semi-infinite-beam solutions still diverge greatly when the response time is greater. For beams with finite extents, multiple reflections must be considered. By applying the hybrid method in the Laplace transform space w.r.t. time, Su and Pao [8] demonstrated that a solution that traces the multiple reflections of transient pulses is more accurate than the modal expansion solution at short response times for Timoshenko beams, yet their derivation and calculation are rather complicated, and they did not connect the wave method and the mode method analytically as Wang et al. did [5]. In addition, Mei and Mace also analyzed monochromatic wave transmission and reflection in coupled beams [3], yet no similar work is implemented for the infinite-beam GF.

This paper tries to find a simple, general method to obtain the exact and short-time-convergent finite-beam GF applicable to any linear boundary conditions by modifying the traditional wave method, and to prove the wave-mode duality of the finite-beam GF analytically. It has been discovered that the reflection and transmission analysis of the self-similar infinite-beam GF can be easily applied in the complex Fourier transform domain (or the complex wavenumber domain for wave equation). The complex-variable self-similar GF for infinite beams is derived. The solution expanded by self-similar functions, which converge better at short response times, is rederived in this paper to cover more general cases. By applying the geometric summation formula for matrices, the superposed infinite-beam GFs are transformed into a new type of modal expansion solutions, which are equivalent to the traditional modal expansion solutions but require no calculation of the integral inner product of each mode. The relations between infinite, semi-infinite, and finite-beam solutions are well connected by reflection analysis.

The evanescent waves in beams, often referred to as 'non-propagating waves', are proved to be 'propagating' (just with imaginary wavenumbers) and equally important as 'propagating waves.' Moreover, the new method can also be applied to other systems, e.g., heat conduction equation. The heat equation GF is also commonly applied in engineering problems (e.g., Fernandes et el. [10]). It is found by the new method that the beam equation possesses dissipative properties, and heat conduction can also be analyzed as propagating waves, just with imaginary angular frequencies.

In brief, this paper filled the gap of the poor convergence of the GF at short times; the newly derived modal solutions are compared with those obtained by the traditional approach; the newly developed method is proved applicable to a batch of linear partial differential equations.

This paper is organized as follows. In section 2, the complex-variable self-similar GF and its series decomposition for infinite beams are deduced. In section 3, the GFs of finite beams are obtained by superimposing reflected self-similar waves, from which the new modal expansion solutions are also derived by applying the geometric summation formula. In section 4, the GF of a coupled finite beam is derived. The new method is applied to the one-dimensional diffusion equation in section 5. The GFs obtained by the traditional and new expansion methods are compared in sections 3 and 5. Summary and conclusions are stated in section 6.

## 2  The governing equation of Euler-Bernoulli beams and the Green's function of infinite beams

The Green's function approach is an elegant and rigorous tool for solving a linear partial differential equation's initial-boundary value problem. Under a concentrated source in time and space, the solution of the differential equation associated with certain initial and boundary conditions, known as a Green's function (GF), is mathematically singular at the source location. When the span of the beam is described along the horizontal $x$ axis as shown in Figure 1, the transverse displacement $w$ along the $z$ axis based on Euler beam theory is governed by:

$$w_{,tt} + c^2 w_{,xxxx} = q \tag{1}$$

where $q$ is the reduced load density in (m·s$^{-2}$), and $c = \sqrt{EI/\mu}$ in (m$^2$·s$^{-1}$), where $E$ is Young's modulus in (Pa), $I$ is the moment of inertia about the $y$ axis in (m$^4$), and $\mu$ is the linear density in (kg·m$^{-1}$). $EI$ is also called the bending rigidity of a beam. The GF $G(x,\xi,t)$ is obtained by solving Eq. (1) with the null initial condition under a unit impulse $\delta(x-\xi)\delta(t)$, where $\delta$ denotes the delta impulse function and $\xi$ refers to the impulse position. If a unit momental impulse is applied, the reduced load becomes $\delta_{,x}(x-\xi)\delta(t)$, and the GF can be derived as $G_{,x}(x,\xi,t)$.

Notice that high-frequency waves in an Euler-Bernoulli beam model possess unbounded phase velocities, which will introduce an unphysical zero-time response. This can be avoided by applying more complicated models (e.g., Timoshenko beam model); nevertheless, the simpler Euler-Bernoulli beam model is applied in this paper to introduce the new method.

Once the GF is obtained, the response of such a beam under an arbitrary load $q(x,t)$ with null initial conditions is:

$$w(x,t) = \int_{-\infty}^{\infty} \int_0^t G(x,\xi,t-\tau) q(\xi,\tau) \, d\tau \, d\xi \tag{2}$$

When waves have not reached the end boundaries, the dynamic response of the beam can be conveniently modeled with an infinite extent, as shown in Figure 1:

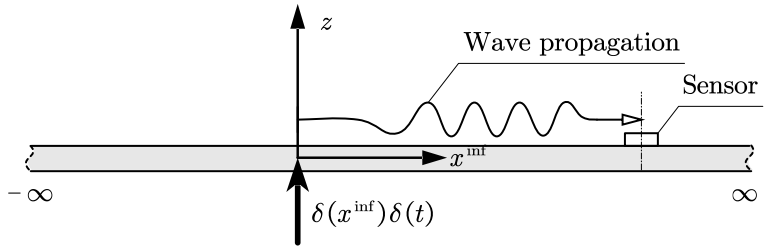

Figure 1. The coordinate system of an infinite beam with a unit force-impulse applied at location $x^{\text{inf}} = \xi = 0$ and time $t = 0$.

Notice that $x^{\text{inf}}$ is specially denoted as the infinite-beam coordinate, as $x$ will be applied later for the finite-beam coordinate. For simplicity, assume the impulse is applied at $x^{\text{inf}} = 0$ and $t = 0$. The GF of an infinite beam $G^{\text{inf}}$ satisfies:

$$G^{\text{inf}}_{,tt} + c^2 G^{\text{inf}}_{,xxxx} = \delta(x^{\text{inf}})\delta(t), \quad -\infty < x < \infty \tag{3}$$

In this paper, the Fourier transform pairs are defined as:

$$\overline{G}(k) = \int_{-\infty}^{\infty} G(x) e^{-ikx} dx$$
$$G(x) = \frac{1}{2\pi} \int_{-\infty}^{\infty} \overline{G}(k) e^{ikx} dk \tag{4}$$

Consider null boundary conditions and null initial conditions. Applying Fourier transform and Laplace transform to Eq. (3) w.r.t. $x^{\text{inf}}$ ($x^{\text{inf}} \to k$) and $t$ ($t \to s$) yields:

$$\overline{\overline{G}}^{\text{inf}}(k,s) = \frac{1}{c^2 k^4 + s^2} \tag{5}$$

Applying inverse Laplace transform to Eq. (5) w.r.t. $s$ yields:

$$\overline{G}^{\text{inf}}(k,t) = \frac{i}{ck^2} e^{-ick^2 t} \tag{6}$$

Different from Graff's work [1], an imaginary part is added to Eq. (6) because the reflection analysis is much easier to conduct exponentially. Applying inverse Fourier transform to Eq. (6) w.r.t. $k$ yields the GF of an infinite beam:

$$G^{\text{inf}}(x^{\text{inf}}, t) = \frac{\sqrt{2\pi ct}}{2c} \mathcal{G}\left(\frac{x^{\text{inf}}}{\sqrt{2\pi ct}}\right) \tag{7}$$

where $\mathcal{G}(\eta)$ is a complex-variable self-similar function defined as:

$$\mathcal{G}(\eta) = \frac{1-i}{\pi} e^{\frac{i\pi\eta^2}{2}} - \eta\, \text{erfi}\left(\frac{1+i}{2}\sqrt{\pi}\eta\right) + i\sqrt{\eta^2} \tag{8}$$

where $\text{erfi}(\eta)$ is the imaginary error function defined as:

$$\operatorname{erfi}(\eta) = -i\operatorname{erf}(i\eta) = \frac{2}{\sqrt{\pi}} \int_0^\eta e^{u^2} du \tag{9}$$

$\mathcal{G}(\eta)$ and the imaginary error function need numerical calculation. $\mathcal{G}(\eta)$ is an even function. Graff derived the self-similar function as a real-variable function, which is in fact $\operatorname{Re}[\mathcal{G}(\eta)]$, as plotted in Figure 2.

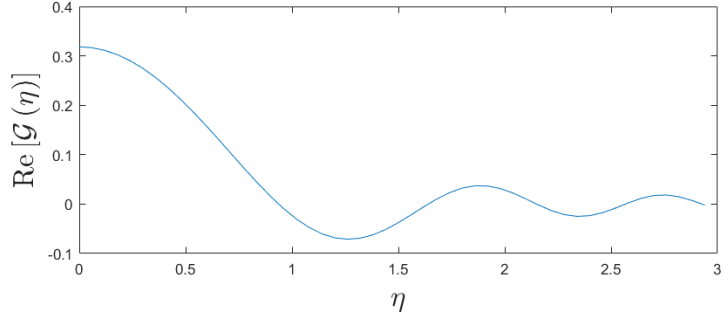

Figure 2. A plot of $\operatorname{Re}[\mathcal{G}(\eta)]$ derived by Graff [1]. $\eta$ in this plot is a real variable. As $\mathcal{G}(\eta) = \mathcal{G}(-\eta)$, the curve is plotted for $x > 0$.

The difference between Eq. (8) and Graff's solution begins from the addition of the complex part in Eq. (6). Series decompositions of $\mathcal{G}(\eta)$ at $\eta = 0$ and $\eta = \infty$ are provided to raise computational efficiency:

$$\mathcal{G}(\eta) = \begin{cases} i\sqrt{\eta^2} - (1+i)\exp(i\pi\eta^2/2) \sum_{n=0}^{N_1} \frac{n!2^n}{(2n)!}(-i\pi)^{n-1} \eta^{2n}, & \operatorname{abs}(\eta) \leqslant \eta_0 \\ -(1+i)\exp(i\pi\eta^2/2) \sum_{n=1}^{N_2} \frac{(2n)!}{n!2^n} \frac{1}{(i\pi)^{n+1} \eta^{2n}}, & \operatorname{abs}(\eta) > \eta_0 \end{cases} \tag{10}$$

where $N_1$, $N_2$ and $\eta_0$ are parameters to set (e.g., $N_1 = 30$, $N_2 = 10$, and $\eta_0 = 2.348$). The validity of Eq. (10) can be easily proved by the equality of the second derivatives of Eq. (8) and (10). Considering $\mathcal{G}(\eta)$ is an even function, the first quadrant of $\mathcal{G}(\eta)$ is plotted in Figure 3:

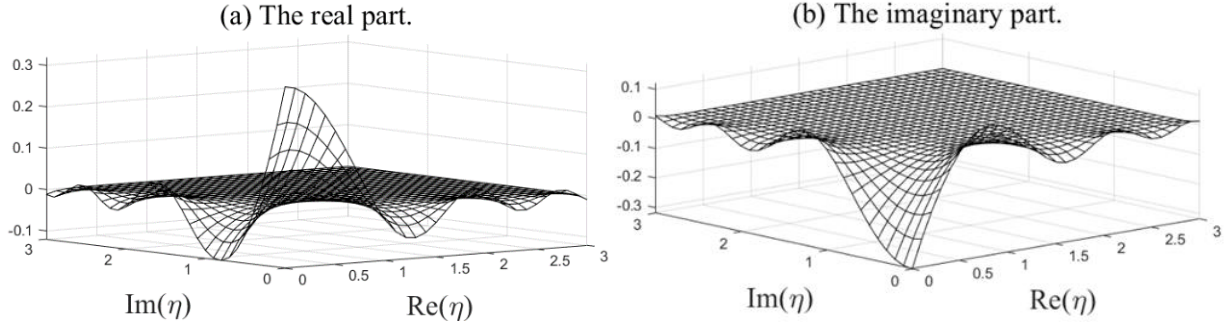

Figure 3. Plots of $\text{Re}(\mathcal{G})$ and $\text{Im}(\mathcal{G})$ for $\text{Re}(\eta) \in [0,3]$ and $\text{Im}(\eta) \in [0,3]$.

$\mathcal{G}(\eta)$ vanishes as $\eta$ approaches infinity in the first and third quadrants of the complex plane, which explains why the infinite-domain GF is accurate enough for short-time responses. The real part and the imaginary part of $\mathcal{G}(\eta)$ are antisymmetric about $\text{Re}[\eta] = \text{Im}[\eta]$. As it will be shown in detail in the following sections, the propagation along the real axis represents 'propagating waves', while the propagation along the imaginary axis represents 'evanescent waves'; the infinite-beam GFs propagating and reflecting along the real axis will compose sinusoidal standing waves, while those along the imaginary axis will compose hyperbolic standing waves. From the antisymmetry of $\mathcal{G}(\eta)$ in Figure 3, propagating and evanescent waves are equivalent except for the propagating direction: parallel to the real axis or the imaginary axis.

The self-similar solution $G^{\text{inf}}(x^{\text{inf}}, t)$ gradually spreads from the origin to infinity as time increases. Considering the dispersion relation $\omega = ck^2$ of a beam is not linear, waves with higher frequencies and greater wavenumbers travel faster than those with not. The solution contains information under every possible frequency and wavenumber.

## 3 Reflection analysis of finite beams

### 3.1 Superposition of self-similar waves in the wavenumber domain

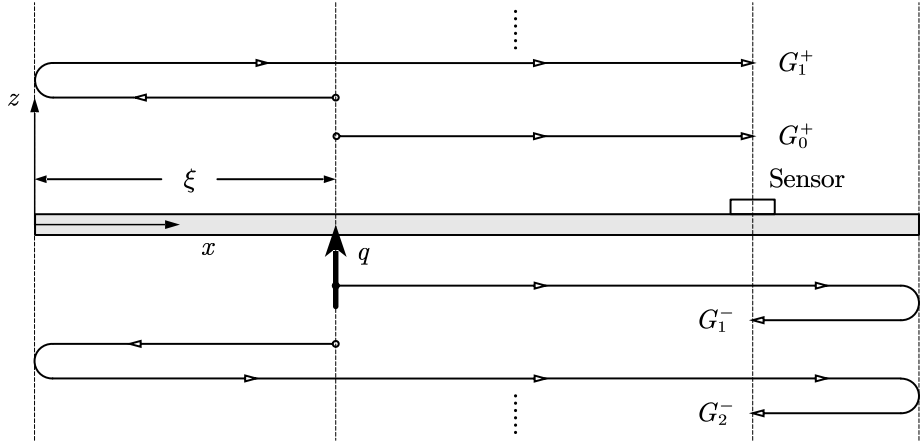

Figure 4. The coordinate system of a finite beam and an intuitive interpretation of $G_n^{\pm}$

A more commonly used coordinate system for finite beams shown in Figure 4 is applied in this section. The origin is set at the left end of the beam. Assume the impulse is exerted at $x=\xi$ and the beam length is $l$. The governing equation becomes:

$$G_{,tt} + c^2 G_{,xxxx} = \delta(x-\xi)\delta(t), \quad 0<x<l \tag{11}$$

Reflection occurs with the existence of boundaries. Reflected waves are denoted as $G_n^{\pm}(x,\xi,t)$, where the superscript $\pm$ denotes the direction of propagation, while the subscript $n$ denotes the times of reflections that waves have encountered. $G_n^{\pm}$ are also intuitively interpreted in Figure 4. The reflected waves can be derived from the infinite-beam GF. The following analysis will be mainly applied in the $k$ wavenumber domain, where $k$ is the Fourier transform variable w.r.t. $x^{\inf}$. From the property of Fourier transform,

$$\mathcal{F}_{x^{\inf} \to k}\left[G^{\inf}(x^{\inf}+x,t)\right] = e^{ikx}\bar{G}^{\inf}(k,t) \tag{12}$$

The reflected waves can be expressed as:

$$\bar{G}_n^+ = \left(a^+ + a_N^+\right)_n \bar{G}^{\inf}(k,t) \qquad \bar{G}_n^- = \left(a^- + a_N^-\right)_n \bar{G}^{\inf}(k,t) \tag{13}$$

where $a^{\pm}$ and $a_N^{\pm}$ are functions of $x$, $\xi$, and $l$, which record the distances that waves have traveled. $a^{\pm}$ represent the propagating waves, while $a_N^{\pm}$ stand for the near-field (or evanescent) waves. It will be convenient to group the wave amplitudes into 2×1 vectors of positive-going waves $\mathbf{a}_n^+$ and negative-going waves $\mathbf{a}_n^-$:

$$\mathbf{a}_n^+ = \begin{bmatrix} a^+ \\ a_N^+ \end{bmatrix}_n, \quad \mathbf{a}_n^- = \begin{bmatrix} a^- \\ a_N^- \end{bmatrix}_n \tag{14}$$

According to Eq. (12), the propagation operator $P$ is defined to denote the distance (designated by its subscript) that a wave has traveled:

$$P_x = e^{ikx}, \quad P_{ix} = e^{-kx} \tag{15}$$

An imaginary subscript of $P$ denotes a propagation state as an evanescent wave. Corresponding to $\mathbf{a}_n^{\pm}$, the propagation matrix $\mathbf{P}_x$ is defined as:

$$\mathbf{P}_x = \begin{bmatrix} P_x & 0 \\ 0 & P_{ix} \end{bmatrix} = \begin{bmatrix} e^{ikx} & 0 \\ 0 & e^{-kx} \end{bmatrix} \tag{16}$$

Direct waves $\mathbf{a}_0^{\pm}$ are naturally derived from the relationship between $G^{\text{inf}}(x^{\text{inf}},t)$, $G_0^+(x,\xi,t) = G^{\text{inf}}(x-\xi,t)$, and $G_0^-(x,\xi,t) = G^{\text{inf}}(\xi-x,t)$:

$$\begin{aligned} \mathbf{a}_0^+ &= \mathbf{P}_{x-\xi}\mathbf{a}_{1,0}^{\text{T}} = \begin{bmatrix} P_{x-\xi} \\ 0 \end{bmatrix} \\ \mathbf{a}_0^- &= \mathbf{P}_{\xi-x}\mathbf{a}_{1,0}^{\text{T}} = \begin{bmatrix} P_{\xi-x} \\ 0 \end{bmatrix} \end{aligned} \tag{17}$$

where $\mathbf{a}_{1,0}^{\text{T}} = \begin{bmatrix} 1 & 0 \end{bmatrix}^{\text{T}}$. However, according to the antisymmetric property of $\mathcal{G}(\eta)$, we know that $\mathcal{G}(\eta) = i\mathcal{G}(i\eta)$. $\mathbf{a}_0^{\pm}$ defined by Eq. (17) is therefore equivalent to:

$$\begin{aligned} \mathbf{a}_0^+ &= \frac{1}{2}\mathbf{P}_{x-\xi}\mathbf{a}_{1,i}^{\text{T}} = \frac{1}{2}\begin{bmatrix} P_{x-\xi} \\ iP_{i(x-\xi)} \end{bmatrix} \\ \mathbf{a}_0^- &= \frac{1}{2}\mathbf{P}_{\xi-x}\mathbf{a}_{1,i}^{\text{T}} = \frac{1}{2}\begin{bmatrix} P_{\xi-x} \\ iP_{i(\xi-x)} \end{bmatrix} \end{aligned} \tag{18}$$

Eq. (18) considers propagating and evanescent waves equally, and it must be applied in the geometric summation formula, as shown in subsection 3.3; while in the other cases, Eq. (17) is applied for simplicity, for it takes half of the number of expressions to reach the same result compared with Eq. (18).

When waves are incident upon a boundary, they give rise to reflected waves and propagate in the opposite direction. The recursive relation between $\mathbf{a}_n^\pm$ and $\mathbf{a}_{n+1}^\pm$ can be expressed as:

$$\begin{aligned} \mathbf{a}_{n+1}^- &= \mathbf{P}_{l-x} \mathbf{R}^+ \mathbf{P}_{l-x} \mathbf{a}_n^+ \\ \mathbf{a}_{n+1}^+ &= \mathbf{P}_x \mathbf{R}^- \mathbf{P}_x \mathbf{a}_n^- \end{aligned} \quad (19)$$

where $\mathbf{R}^+$ and $\mathbf{R}^-$ are reflection matrices of the right and left boundary, respectively. By letting $\mathbf{a}_n^\pm + \mathbf{a}_{n+1}^\mp$ satisfy the boundary conditions, the reflection matrix of a boundary that connects translational and rotational springs with spring constants $k_t$ and $k_r$, and translational and rotational damping ratios denoted by $c_t$ and $c_r$, and a mass $m$ with a moment of inertia $I_m$, is obtained:

$$\mathbf{R}^g = \begin{bmatrix} k_t - ic_t\omega - m\omega^2 - iEIk^3 & k_t - ic_t\omega - m\omega^2 - EIk^3 \\ -ik_r - c_r\omega + iI_m\omega^2 - EIk & k_r - ic_r\omega - I_m\omega^2 + EIk \end{bmatrix}^{-1} \begin{bmatrix} -k_t + ic_t\omega + m\omega^2 - iEIk^3 & -k_t + ic_t\omega + m\omega^2 - EIk^3 \\ -ik_r - c_r\omega + iI_m\omega^2 + EIk & k_r - ic_r\omega - I_m\omega^2 - EIk \end{bmatrix}$$

(20)

If only the effects of springs are considered, the reflection matrix of the elastically clamped boundary is:

$$\mathbf{R}^{ec} = \begin{bmatrix} k_t - iEIk^3 & k_t - EIk^3 \\ -ik_r - EIk & k_r + EIk \end{bmatrix}^{-1} \begin{bmatrix} -k_t - iEIk^3 & -k_t - EIk^3 \\ -ik_r + EIk & k_r - EIk \end{bmatrix} \quad (21)$$

The reflection matrices of pinned, clamped, free, and sliding boundaries are degenerated from Eq. (21) when $k_t$ and $k_r$ are equal 0 or $\infty$:

$$\mathbf{R}^p = -\mathbf{I}, \quad \mathbf{R}^c = -\begin{bmatrix} -i & 1-i \\ 1+i & i \end{bmatrix}, \quad \mathbf{R}^f = \begin{bmatrix} i & 1-i \\ 1+i & -i \end{bmatrix}, \quad \mathbf{R}^s = \mathbf{I} \quad (22)$$

where $\mathbf{I}$ is the 2×2 identity matrix. Reflection matrices apply to both the right and the left boundaries.

The general formulae of $\mathbf{a}_n^{\pm}$ are obtained from the recursive relation Eq. (19):

$$\mathbf{a}_n^+ = \begin{cases} \mathbf{P}_x \mathbf{R}^- \mathbf{P}_l \left( \mathbf{R}^+ \mathbf{P}_l \mathbf{R}^- \mathbf{P}_l \right)^{\frac{n-2}{2}} \mathbf{R}^+ \mathbf{P}_{l-x} \mathbf{a}_0^+, & n \text{ is even}, n \in \mathbb{N}^+ \\ \mathbf{P}_x \mathbf{R}^- \mathbf{P}_l \left( \mathbf{R}^+ \mathbf{P}_l \mathbf{R}^- \mathbf{P}_l \right)^{\frac{n-1}{2}} \mathbf{P}_{x-l} \mathbf{a}_0^-, & n \text{ is odd}, n \in \mathbb{N}^+ \end{cases} \quad (23)$$

$$\mathbf{a}_n^- = \begin{cases} \mathbf{P}_{l-x} \left( \mathbf{R}^+ \mathbf{P}_l \mathbf{R}^- \mathbf{P}_l \right)^{\frac{n-1}{2}} \mathbf{R}^+ \mathbf{P}_{l-x} \mathbf{a}_0^+, & n \text{ is odd}, n \in \mathbb{N} \\ \mathbf{P}_{l-x} \left( \mathbf{R}^+ \mathbf{P}_l \mathbf{R}^- \mathbf{P}_l \right)^{\frac{n}{2}} \mathbf{P}_{x-l} \mathbf{a}_0^-, & n \text{ is even}, n \in \mathbb{N} \end{cases} \quad (24)$$

The direct wave is considered as $\mathbf{a}_0^-$ in Eq. (24), and therefore, it is not repeated as $\mathbf{a}_0^+$ in Eq. (23), considering $G^{\text{inf}}$ is an even function. Summing up all of the terms in Eq. (23) and Eq. (24) yields the superimposed amplitude, which by multiplying $G^{\text{inf}}(k,t)$ and applying inverse Fourier transform yields the GF in the spatial domain:

$$G(x,\xi,t) = \left\{ \mathcal{F}_{k \to x^{\text{inf}}}^{-1} \left[ \mathbf{a}_{1,1} \left( \mathbf{P}_{-x} + \mathbf{P}_x \mathbf{R}^- \right) \mathbf{P}_l \sum_{n=0}^{\infty} \left( \mathbf{R}^+ \mathbf{P}_l \mathbf{R}^- \mathbf{P}_l \right)^n \left( \mathbf{R}^+ \mathbf{P}_{l-x} \mathbf{a}_0^+ + \mathbf{P}_{x-l} \mathbf{a}_0^- \right) \overline{G}^{\text{inf}}(k,t) \right] \right\}_{x^{\text{inf}}=0}$$

(25)

where $\mathbf{a}_{1,0} = [1 \ 0]$. The propagation operators already denote the effects of propagation; therefore, $x^{\text{inf}} = 0$ is applied after the inverse transform. Equations before Eq. (25) in this subsection are consistent with previous works (e.g., [2], [3]) but in the wavenumber domain, while Eq. (25) is a new attempt to sum up self-similar waves in the wavenumber domain.

As a simple example, for a simply supported beam, substituting Eq. (17) and $\mathbf{R}^{\pm} = -\mathbf{I}$ into Eq. (25) yields:

$$G(x,\xi,t) = \sum_{n=-\infty}^{\infty} \left[ G^{\text{inf}}(x-\xi+2nl,t) - G^{\text{inf}}(x+\xi+2nl,t) \right] \quad (26)$$

Eq. (25) becomes the GF of a semi-infinite beam lying along $x > 0$ when $\mathbf{R}^+ = \mathbf{0}$:

$$G(x,\xi,t) = G^{\text{inf}}(x-\xi,t) + \mathcal{F}^{-1}_{k\to\xi+x}\left[\text{R}^-_{11}\bar{G}^{\text{inf}}(k,t)\right] + \mathcal{F}^{-1}_{k\to\xi+ix}\left[\text{R}^-_{21}\bar{G}^{\text{inf}}(k,t)\right] \quad (27)$$

where $\text{R}^-_{11}$ and $\text{R}^-_{21}$ are the coefficients of the reflection matrix $\mathbf{R}^-$ for a given boundary condition at $x=0$. Eq. (27) is derived from Eq. (17) rather than Eq. (18), otherwise more terms will appear in Eq. (27). Eq. (27) is a general expression applicable to any semi-infinite beams with linear boundary conditions, thus it is a generalization of Akkaya T. and Van Horssen's work [7]. For semi-infinite beams with four classical boundary conditions, the GFs are:

$$\underbrace{G(x,\xi,t)}_{\text{semi-infinite beams}} = \begin{cases} \text{Pinned}: & G^{\text{inf}}(x-\xi,t) - G^{\text{inf}}(\xi+x,t) \\ \text{Sliding}: & G^{\text{inf}}(x-\xi,t) + G^{\text{inf}}(\xi+x,t) \\ \text{Clamped}: & G^{\text{inf}}(x-\xi,t) + iG^{\text{inf}}(\xi+x,t) - (1+i)G^{\text{inf}}(\xi+ix,t) \\ \text{Free}: & G^{\text{inf}}(x-\xi,t) + iG^{\text{inf}}(\xi+x,t) + (1+i)G^{\text{inf}}(\xi+ix,t) \end{cases} \quad (28)$$

## 3.2 Comparison between the results calculated by two different expansion methods

The most widely applied analytical approach to obtain the beam response is the traditional modal expansion method. The GF expressed by the traditional modal analysis is:

$$G(x,\xi,t) = \sum_n \frac{X_n(\xi)X_n(x)}{\langle X_n(x), X_n(x)\rangle} \bar{G}^{\text{inf}}(k_n,t) \quad (29)$$

where $X_n(x)$ is the $n^{\text{th}}$-order normal mode determined by specific boundary conditions, and $\langle X_m, X_n\rangle$ is the inner product between two modes, which for a single-span beam can be expressed as:

$$\langle X_m, X_n\rangle = \int_0^l X_m X_n dx + \frac{1}{k_m^2 - k_n^2}\left[X_m X_{n,xxx} - X_{m,xxx}X_n - X_{m,x}X_{n,xx} + X_{m,xx}X_{n,x}\right]_0^l \quad (30)$$

It is obvious that $\langle X_m, X_n\rangle = 0$ if $m \neq n$. Eqs. (29) and (30) can be found in any structural dynamics textbook (e.g., [1] and [4]).

The new approach described in section 3.1 and calculated by Eq. (25) is compared with the traditional approach calculated by Eq. (29). For simplicity, define dimensionless time $t^*$:

$$t^* = t/\tau_l, \quad \tau_l = l^2/c \quad (31)$$

Assume a beam is simply supported, and that $\xi = 0.5l$, $x = 0.7l$, and $0 \leqslant t^* \leqslant 0.3$. We do not need to set parameter $c$ if we nondimensionalize $G$ as $G^*$:

$$G^* = Gl/\tau_l \tag{32}$$

In Figure 5, the summations of the first eight terms of each approach are plotted. The self-similar approach shows more details at short times, while it loses details at long times; the traditional expansion solution seems to keep the same scale of accuracy all the time. Two solutions superpose each other at $t^* \approx 0.08$.

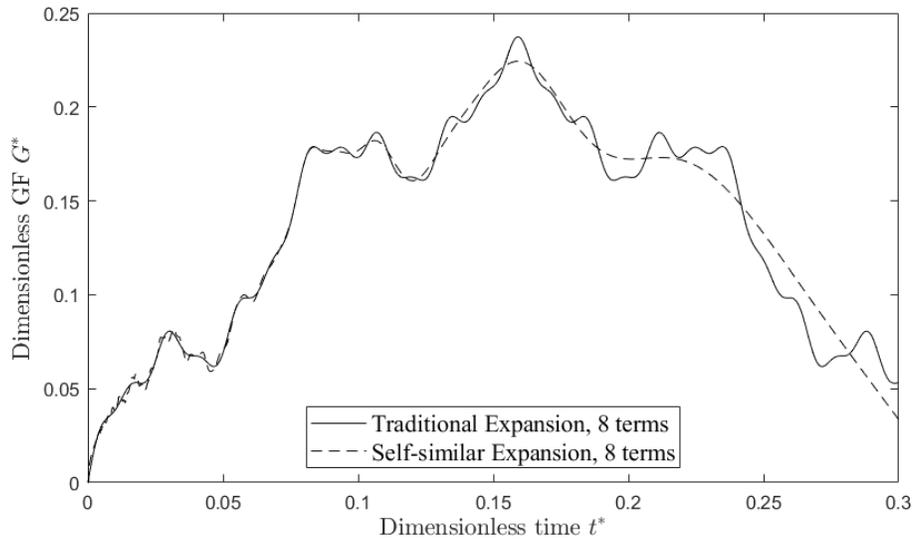

Figure 5. Comparison of two types of solutions for a simply supported beam.

To further show that solutions obtained from both methods converge to the same solution, solutions obtained with the same parameters are plotted in Figure 6, except that the first 16 terms are used instead. Still, the two solutions coincide well around $t^* \approx 0.8$.

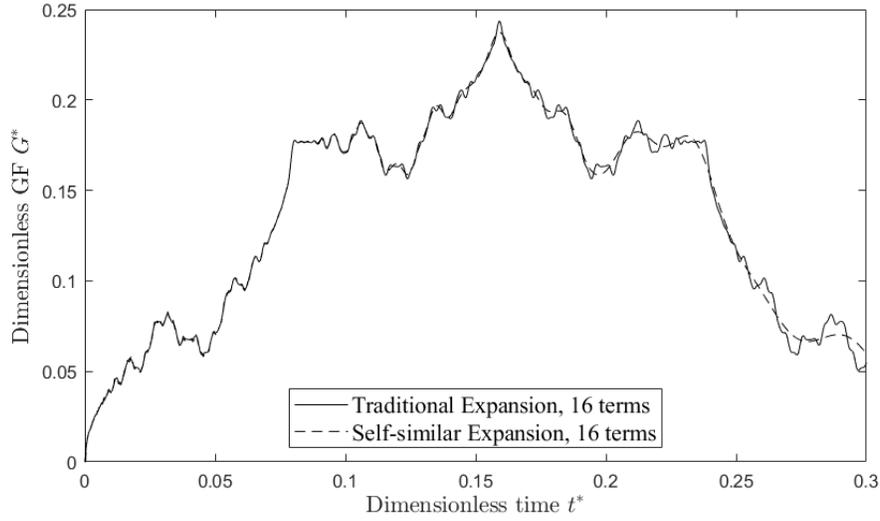

Figure 6. Comparison of two types of solutions for a simply supported beam at long times.

The convergence rate of the two approaches can be roughly compared. Taking a simply supported beam for example, for given location $x$ and time $t$, the decaying property of Eq. (25) is given by the exponential term in $\mathcal{G}(\eta)$, and the leading term is $\exp(-n^2 l^2 / ct)$. The exponential term must decay fast for large $n$, however, the decaying rate will decrease as $t$ increases. The converging rate of the traditional modal expansion method is about $c/(\omega_n l^2) = c/(n\pi)^2$, which is indeed time-independent. A reference threshold for 'the short time' and 'the long time' obtained from Figure 5 and Figure 6 is $t^* \approx 0.08$, i.e., with the same location $x$ and the number of terms $n$, the self-similar expansion approach is suggested when $t^* < 0.08$, otherwise the traditional modal expansion approach is suggested to reach higher accuracy.

The properties of two types of expansion methods are qualitatively summarized in Table 1:

Table 1. Qualitative Comparison of Two Methods

|  | **Self-similar wave expansion** | **Traditional modal expansion** |
|---|---|---|
| **Expanded Functions** | Self-similar functions | Orthogonal modal functions |

| Initial Condition / Impulse Approximation | The first term satisfies initial conditions perfectly, while the rest 'reflected' terms are zero at the beginning | Asymptotically satisfied by superimposing more modal terms |
|---|---|---|
| **Boundary Conditions** | Asymptotically satisfied by superimposing more reflected terms | Each term satisfies boundary conditions |
| **Accuracy/Efficiency** | More accurate/efficient at short times | More accurate/efficient at long times |

### 3.3 Application of the geometric summation formulas for matrix series

Eq. (25) can be transformed with the geometric summation formula for matrix series, which for an arbitrary matrix $\mathbf{M}$ can be expressed as ([8]):

$$\sum_{n=0}^{\infty} \mathbf{M}^n = (\mathbf{I} - \mathbf{M})^{-1} \tag{33}$$

Eq. (33) can be proved by multiplying $(\mathbf{I} - \mathbf{M})$ to both sides. Substituting Eq. (18) and Eq. (33) with $\mathbf{M} = \mathbf{R}^+ \mathbf{P}_l \mathbf{R}^- \mathbf{P}_l$ into Eq. (25) yields:

$$G(x,\xi,t) = \frac{1}{2} \left[ \mathcal{F}_{k \to x^{\inf}}^{-1} \left[ \mathbf{M}_1 \mathbf{M}_2^{-1} \mathbf{M}_3 \bar{G}^{\inf}(k,t) \right] \right]_{x^{\inf}=0} \tag{34}$$

where:

$$\begin{aligned}
\mathbf{M}_1 &= \mathbf{a}_{1,1} \left( \mathbf{P}_{-x} + \mathbf{P}_x \mathbf{R}^- \right) \\
\mathbf{M}_2 &= \left( \mathbf{I} - \mathbf{R}^+ \mathbf{P}_l \mathbf{R}^- \mathbf{P}_l \right) \mathbf{P}_{-l} = \mathbf{P}_{-l} - \mathbf{R}^+ \mathbf{P}_l \mathbf{R}^- \\
\mathbf{M}_3 &= \left( \mathbf{R}^+ \mathbf{P}_{l-\xi} + \mathbf{P}_{\xi-l} \right) \mathbf{a}_{1,i}^{\mathrm{T}}
\end{aligned} \tag{35}$$

Simplified expressions of $\mathbf{M}_1$ and $\mathbf{M}_3$ for four classical boundaries are:

$$\mathbf{M}_1 = \begin{cases} \text{Pinned:} & 2\left[-i\sin(kx) \quad \sinh(kx)\right] \\ \text{Clamped:} & 2(1+i)\left[\phi_4(kx)-\phi_3(kx) \quad \phi_4(kx)-i\phi_3(kx)\right] \\ \text{Free:} & 2(1+i)\left[\phi_1(kx)-\phi_2(kx) \quad -i\phi_1(kx)+\phi_2(kx)\right] \\ \text{Sliding:} & 2\left[\cos(kx) \quad \cosh(kx)\right] \end{cases} \tag{36}$$

$$\mathbf{M}_3 = \begin{cases} \text{Pinned:} & 2i\left[-\sin(k(l-\xi)) \quad \sinh(k(l-\xi))\right]^{\mathrm{T}} \\ \text{Clamped:} & 2(1+i)\left[\phi_4(k(l-\xi))-\phi_3(k(l-\xi)) \quad \phi_3(k(l-\xi))+i\phi_4(k(l-\xi))\right]^{\mathrm{T}} \\ \text{Free:} & 2(1+i)\left[\phi_1(k(l-\xi))-\phi_2(k(l-\xi)) \quad \phi_1(k(l-\xi))+i\phi_2(k(l-\xi))\right]^{\mathrm{T}} \\ \text{Sliding:} & 2\left[\cos(k(l-\xi)) \quad i\cosh(k(l-\xi))\right]^{\mathrm{T}} \end{cases} \tag{37}$$

where $\phi_{1-4}$ are Krylov-Duncan functions defined as:

$$\begin{Bmatrix} \phi_1(x) \\ \phi_2(x) \\ \phi_3(x) \\ \phi_4(x) \end{Bmatrix} = \frac{1}{2} \begin{Bmatrix} \cosh x + \cos x \\ \sinh x + \sin x \\ \cosh x - \cos x \\ \sinh x - \sin x \end{Bmatrix} \tag{38}$$

The inverse of $\mathbf{M}_2$ is:

$$\mathbf{M}_2^{-1} = \frac{\mathbf{M}_2^*}{\det(\mathbf{M}_2)} \tag{39}$$

where * means taking the adjugate matrix. Calculating $\det(\mathbf{M}_2)$ yields the characteristic function $\mathcal{F}(kl)$, whose roots are the characteristic wavenumbers $k_n$ which also determine the natural frequencies $\omega_n = ck_n^2$.

By defining $z = kl$, at each root $z_n$, the characteristic function $\mathcal{F}(z)$ crosses the $z$ axis. The reciprocal of $\mathcal{F}(z)$ gives rise to imaginary impulse functions. To prove it, applying Taylor Series Expansion of $\mathcal{F}(z)$ at $z = z_n$ and the residue theorem of a simple pole on the real axis to the L.H.S. of Eq. (40) yields:

$$\int_{z_n-\varepsilon}^{z_n+\varepsilon} \frac{\mathcal{F}_{,z}(z_n)}{\mathcal{F}(z)} \mathrm{d}z = \int_{z_n-\varepsilon}^{z_n+\varepsilon} \frac{1}{z-z_n} \mathrm{d}z = -i\pi \tag{40}$$

$\text{Im}[1/(z-z_n)] = 0$ along the real axis except at $z = z_n$. Therefore, taking the derivative of Eq. (40) yields impulse functions:

$$\frac{\mathcal{F}_{,z}(z_n)}{\mathcal{F}(z)} = -i\pi \sum_n \delta(z - z_n) \tag{41}$$

The property of the delta impulse function converts Eq. (34) into a superposition of specific vibrational modes, each under a corresponding characteristic wavenumber $k_n$. Eq. (34) can then be rewritten as:

$$G(x,\xi,t) = \sum_{n=-\infty}^{\infty} \frac{\mathbf{M}_1 \mathbf{M}_2^* \mathbf{M}_3}{2i \det(\mathbf{M}_2)_{,k}(k_n)} \bar{G}^{\text{inf}}(k_n, t) \tag{42}$$

Eq. (42) is a superposition of vibrational modes. It is a new type of normal-mode solution requiring no inner-product calculation.

### 3.4 Specific cases for the new modal-expansion GF

For a simply supported beam,

$$\begin{aligned} G(x,\xi,t) &= \sum_{n=1}^{\infty} \frac{i e^{-i\omega_n t}}{\omega_n l} \left[ \sin(k_n x) \sin(k_n \xi) - \sinh(i k_n x) \sinh(i k_n \xi) \right] \\ &= \sum_{n=1}^{\infty} \frac{2i e^{-i\omega_n t}}{\omega_n l} \sin(k_n x) \sin(k_n \xi) \end{aligned} \tag{43}$$

where $k_n = n\pi/l$ and $\omega_n = c k_n^2$. For a clamped-pinned beam,

$$G(x,\xi,t) = \sum_{n=1}^{\infty} \frac{4i e^{-i\omega_n t}/\omega_n l}{c_n - c_n^{-1}} \left\{ \begin{array}{l} \sinh(k_n(l-\xi))\left[\cos(k_n l)\phi_4(k_n x) - \sin(k_n l)\phi_3(k_n x)\right] + \\ \sin(k_n(l-\xi))\left[\sinh(k_n l)\phi_3(k_n x) - \cosh(k_n l)\phi_4(k_n x)\right] \end{array} \right\} \tag{44}$$

where $c_n = \cos(k_n l)/\cosh(k_n l)$, and $k_n$ are the positive roots of:

$$\mathcal{F}(kl) = \cos(kl)\sinh(kl) - \sin(kl)\cosh(kl) \tag{45}$$

For a cantilever (clamped-free) beam,

$$G = \sum_{n=1}^{\infty} \frac{8i e^{-i\omega_n t}/\omega_n l}{\sin(k_n l)\cosh(k_n l) + \tanh(k_n l)} \left\{ \begin{array}{l} \phi_1(k_n(l-\xi))\left[\phi_2(k_n l)\phi_3(k_n x) - \phi_1(k_n l)\phi_4(k_n x)\right] \\ +\phi_2(k_n(l-\xi))\left[\phi_4(k_n l)\phi_4(k_n x) - \phi_1(k_n l)\phi_3(k_n x)\right] \end{array} \right\} \tag{46}$$

where $k_n$ are the positive roots of $\mathcal{F}(kl) = \cos(kl)\cosh(kl) + 1$.

## 3.5 Verification of the newly derived modal expansion solution by comparing with the traditional modal expansion solution

Still set $\xi = 0.5l$ and $x = 0.7l$. As we are comparing two modal expansion results, which converge better at long times, we set a relatively long time range $0 \leqslant t^* \leqslant 1$. A comparison of results obtained for a cantilever beam is shown in Figure 7. The high degree of consistency between results shown in Figure 7 verifies the equivalence of Eq. (42) and Eq. (29). In addition, compared with Eq. (29), Eq. (42) avoids calculating the inner product of different modes; thus, the new modal expansion approach is more efficient than the traditional modal expansion solution.

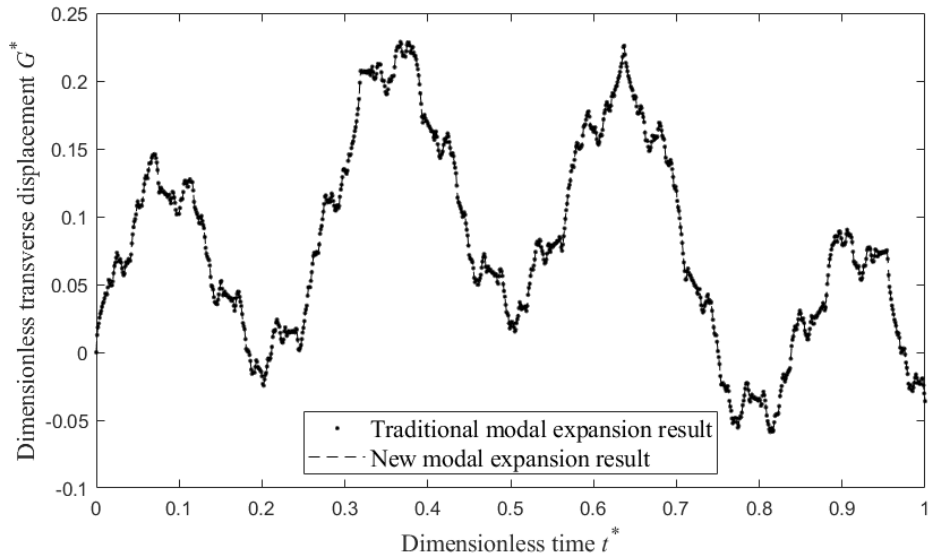

Figure 7. A comparison of the traditional and new modal expansion results for a cantilever beam (first 13 modes for both)

## 4 Reflection analysis of coupled beams

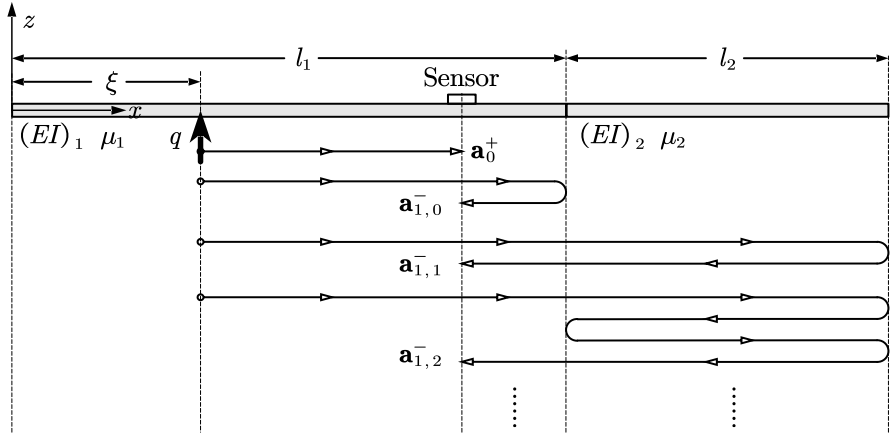

Figure 8. The coordinate system of a coupled beam.

As shown in Figure 8, suppose a coupled beam has subspan lengths $l_1$ and $l_2$, and dispersive parameters $a_1$ and $a_2$, for the left and right span, respectively. An impulse is applied and then detected by a sensor at the left span. The analysis is analogous to that in section 3. The difference comes from the reflection and transmission at the junction at $x = l_1$. Denote $\mathbf{R}_1^-$ and $\mathbf{R}_1^+$ as the reflection matrix of the left and right ends of the left span, respectively, $\mathbf{R}_2^\pm$ for the right span, and $\mathbf{T}^+$ and $\mathbf{T}^-$ as the rightward and leftward transmission matrix at the junction, respectively. $\mathbf{R}_{1,2}^\pm$ and $\mathbf{T}^\pm$ are determined by the boundary conditions and the continuity conditions. For each span, a dispersive relation can be defined:

$$\text{Left}: \omega_1 = c_1 k_1^2 \quad \text{Right}: \omega_2 = c_2 k_2^2 \tag{47}$$

To apply reflection and transmission analysis in the wavenumber domain, assume that $\omega_1 = \omega_2 = \omega$, and denote $k_1 = k$ and $k_2 = rk$, where $r^2 = c_1 / c_2$. Therefore, except for the original propagation matrix $\mathbf{P}_x$ for the left span, a new propagation matrix $\mathbf{P}_{rx}$ should be applied to describe wave propagation on the right span.

Suppose two spans are perfectly connected, i.e., the displacement, slope, moment, and shear force are all continuous at the connection, and for simplicity, denote $\gamma = (EI)_2/(EI)_1$ as the bending rigidity ratio, then the reflection matrix $\mathbf{R}_1^+$ and transmission matrix $\mathbf{T}^+$ at the connection can be obtained as:

$$\mathbf{R}_1^+(\gamma,r) = -\begin{bmatrix} i-r-ir^2\gamma+r^3\gamma & 1+r+r^2\gamma+r^3\gamma \\ 1+r+r^2\gamma+r^3\gamma & -i-r+ir^2\gamma+r^3\gamma \end{bmatrix}^{-1} \begin{bmatrix} -i-r+ir^2\gamma+r^3\gamma & -1+r-r^2\gamma+r^3\gamma \\ -1+r-r^2\gamma+r^3\gamma & i-r-ir^2\gamma+r^3\gamma \end{bmatrix}$$

$$\mathbf{T}^+(\gamma,r) = 4\begin{bmatrix} 1+r+\gamma r^2+\gamma r^3 & 1+ir-\gamma r^2-i\gamma r^3 \\ 1-ir-\gamma r^2+i\gamma r^3 & 1+r+\gamma r^2+\gamma r^3 \end{bmatrix}^{-1}$$

(48)

$\mathbf{R}_2^-$ and $\mathbf{T}^-$ can be obtained by substituting $1/\gamma$ and $1/r$ into $\mathbf{R}_1^+$ and $\mathbf{T}^+$, i.e.,

$$\begin{aligned} \mathbf{R}_2^- &= \mathbf{R}_1^+(1/\gamma,1/r) \\ \mathbf{T}^- &= \mathbf{T}^+(1/\gamma,1/r) \end{aligned}$$

(49)

If the coupled beam is not connected in the way shown in Figure 8, different reflection and transmission matrices can be obtained regarding specific continuity conditions.

On the left span, when a right-going wave $\mathbf{a}_n^+$ meets the junction, some part of it reflects, and the other part propagates through the junction rightward, which will eventually go through the junction again leftward. For simplicity, denote $\mathbf{a}_{n+1}^-$ as the superposition of such two parts. As shown in Figure 8, taking the incidence of $\mathbf{a}_0^+$ as an example, denote $\mathbf{a}_{1,m}^-$ as the reflected amplitude that has propagated over the right beam span and returned to the junction for $m$ times, then,

$$\mathbf{a}_{1,m}^- = \begin{cases} \mathbf{P}_{l_1-x}\mathbf{R}_1^+\mathbf{P}_{l_1-x}\mathbf{a}_0^+, & m=0 \\ \mathbf{P}_{l_1-x}\mathbf{T}^-\mathbf{P}_{rl_2}\mathbf{R}_2^+\mathbf{P}_{rl_2}\left(\mathbf{R}_2^-\mathbf{P}_{rl_2}\mathbf{R}_2^+\mathbf{P}_{rl_2}\right)^{m-1}\mathbf{T}^+\mathbf{P}_{l_1-x}\mathbf{a}_0^+, & m\geqslant 1 \end{cases}$$

(50)

from which

$$\mathbf{a}_1^- = \sum_{m=0}^{\infty}\mathbf{a}_{1,m}^- = \mathbf{P}_{l_1-x}\mathbf{R}^+\mathbf{P}_{l_1-x}\mathbf{a}_0^+$$

(51)

where

$$\mathbf{R}^+ = \mathbf{R}_1^+ + \mathbf{T}^- \mathbf{P}_{rl_2} \mathbf{R}_2^+ \mathbf{P}_{rl_2} \left(\mathbf{I} - \mathbf{R}_2^- \mathbf{P}_{rl_2} \mathbf{R}_2^+ \mathbf{P}_{rl_2}\right)^{-1} \mathbf{T}^+ \qquad (52)$$

The recursive relation between $\mathbf{a}_n^-$ and $\mathbf{a}_{n+1}^+$ keeps the same as that defined in Eq. (19). Therefore, the GF can be derived and expressed similarly as in section 3:

$$G(x,\xi,t) = \frac{1}{2}\left\{\mathcal{F}_{k\to x^{\inf}}^{-1}\left[\mathbf{a}_{1,1}\left(\mathbf{a}_0 + \sum_{n=1}^{\infty}\left(\mathbf{a}_n^+ + \sum_{m=0}^{\infty}\mathbf{a}_{n,m}^-\right)\right)\bar{G}^{\inf}(k,t)\right]\right\}_{x^{\inf}=0} \qquad (53)$$

or if we apply the geometric summation formula twice, Eq. (53) becomes:

$$G(x,\xi,t) = \sum_n \frac{\mathbf{M}_1 \mathbf{M}_2^* \mathbf{M}_3}{2i \det(\mathbf{M}_2)_{,k}(k_n)} \bar{G}^{\inf}(k_n,t) \qquad (54)$$

where

$$\begin{aligned}
\mathbf{M}_1 &= \mathbf{a}_{1,1}\left(\mathbf{P}_{-x} + \mathbf{P}_x \mathbf{R}_1^-\right) \\
\mathbf{M}_2 &= \left(\mathbf{P}_{-rl_2}\left(\mathbf{R}_2^+\right)^{-1}\mathbf{P}_{-rl_2} - \mathbf{R}_2^-\right)\left(\mathbf{T}^-\right)^{-1}\left(\mathbf{P}_{-l_1} - \mathbf{R}_1^+ \mathbf{P}_{l_1}\mathbf{R}_1^-\right) - \mathbf{T}^+ \mathbf{P}_{l_1}\mathbf{R}_1^- \\
\mathbf{M}_3 &= \left[\left(\mathbf{P}_{-rl_2}\left(\mathbf{R}_2^+\right)^{-1}\mathbf{P}_{-rl_2} - \mathbf{R}_2^-\right)\left(\mathbf{T}^-\right)^{-1}\left(\mathbf{P}_{\xi-l_1} + \mathbf{R}_1^+ \mathbf{P}_{l_1-\xi}\right) + \mathbf{T}^+ \mathbf{P}_{l_1-\xi}\right]\mathbf{a}_{1,i}^{\mathrm{T}}
\end{aligned} \qquad (55)$$

The coupled solution degenerates into the one-span case when $r = \gamma = 1$, which yields $\mathbf{R}_1^+ = \mathbf{R}_2^- = \mathbf{0}$ and $\mathbf{T}^\pm = \mathbf{I}$. The coupled solution can be generalized into the three-span case by substituting $\mathbf{R}_2^+$ with an analog of Eq. (52).

As a simple verification, suppose a coupled beam is simply supported at two boundaries, $l_1 = l_2 = \pi/2$ (m), and $c_1 = 1$ (m²/s), $\gamma = 1$, $r = 1.2$. $k_n$ calculated by reflection analysis for the first three modes are 0.89683, 1.8363, and 2.7047 (m⁻¹), while the traditional modal analysis yields the same results.

## 5 Reflection analysis of the one-dimensional heat conduction equation

### 5.1 The infinite-domain solution

The method of reflection and transmission analysis in the Fourier transform domain can also be applied to other types of equations, e.g., heat equation. The GF of the one-dimensional heat conduction equation is solved as an example. The GF is physically equivalent to the short-time temperature distribution of an object under a unit laser pulse. At short times when boundary effects can be neglected, set the center of coordinate at $x^{\text{inf}}=0$ (similar to Figure 1), then the governing equation becomes:

$$T^{\text{inf}}_{,t} - cT^{\text{inf}}_{,xx} = \delta(x^{\text{inf}})\delta(t) \tag{56}$$

where $T$ is the Green's function in [m$^{-1}$], and $c$ is again a constant in (m$^2$/s). This constant is expressed with the same symbol as that defined in the Eq. (1), as the two governing equations share similar dissipating properties. The dispersive property of beams is another type of dissipation, and in the following contents, it will be shown that heat transfer can also be analyzed as propagating waves and $c$ denotes the dissipation rate for both systems.

Applying Fourier transform and Laplace transform to Eq. (56) w.r.t. $x^{\text{inf}}$ and $t$ yields:

$$T^{\text{inf}}(k,s) = 1/(s+ck^2) \tag{57}$$

Applying inverse Laplace transform to Eq. (57) w.r.t. $s$ yields:

$$T^{\text{inf}}(k,t) = e^{-ck^2 t} \tag{58}$$

Applying inverse Fourier transform to Eq. (58) w.r.t. $k$ yields:

$$T^{\text{inf}}(x^{\text{inf}},t) = \frac{1}{\sqrt{4\pi ct}} \exp\left(-\left(x^{\text{inf}}\right)^2/(4ct)\right) \tag{59}$$

$T^{\text{inf}}(x^{\text{inf}},t)$ is the Gaussian distribution w.r.t. $x^{\text{inf}}$ with a variance $\sigma = 2\pi t$. Eq. (59) is usually called the fundamental solution to the heat equation. Its derivation can also be found in any heat transfer textbook.

### 5.2 Finite-domain solutions derived from the infinite-domain solution by reflection analysis

Similar to the coordinate system shown in Figure 4, suppose the heat impulse is applied at $x=\xi$. Only the unsteady part of the response is concerned here, and thus the first, second, and third kinds of homogeneous boundary conditions can be expressed as:

$$1^{\text{st}}: T=0 \quad 2^{\text{nd}}: T_{,x}=0 \quad 3^{\text{rd}}: T_{,x}+bT=0 \tag{60}$$

where $b$ is a constant. Reflected terms are denoted as $T_n^{\pm}(x,\xi,t)$, where the superscript $\pm$ denotes the direction of 'propagation', while the subscript $n$ denotes the times of reflections. The reflected waves can be expressed as:

$$\begin{aligned} \overline{T}_n^+ &= a_n^+ \overline{T}^{\text{inf}}(k,t) \\ \overline{T}_n^- &= a_n^- \overline{T}^{\text{inf}}(k,t) \end{aligned} \tag{61}$$

The propagation operator $P_x = e^{ikx}$ and the direct terms $a_0^{\pm}$ are defined similarly to those in section 3:

$$a_0^+ = P_{x-\xi}, \quad a_0^- = P_{\xi-x} \tag{62}$$

The recursive relation between $a_n^{\pm}$ and $a_{n+1}^{\pm}$ can be expressed as:

$$\begin{aligned} a_{n+1}^- &= P_{l-x} R^+ P_{l-x} a_n^+ \\ a_{n+1}^+ &= P_x R^- P_x a_n^- \end{aligned} \tag{63}$$

where $R^+$ and $R^-$ are reflection coefficients of the right and left boundary, respectively. By letting $a_n^{\pm} + a_{n+1}^{\mp}$ satisfy the boundary conditions, the reflection coefficients of the three types of boundary conditions are obtained:

$$1^{\text{st}}: R = -1 \quad 2^{\text{nd}}: R = 1 \quad 3^{\text{rd}}: R = \frac{ik+b}{ik-b} \tag{64}$$

$a_n^{\pm}$ are obtained from $a_0^{\pm}$ defined by Eq. (62) and the recursive relation Eq. (63):

$$a_n^+ = \begin{cases} P_x R^- P_l \left(R^+ P_l R^- P_l\right)^{\frac{n-2}{2}} R^+ P_{l-\xi}, & n \text{ is even} \\ P_x R^- P_l \left(R^+ P_l R^- P_l\right)^{\frac{n-1}{2}} P_{\xi-l}, & n \text{ is odd} \end{cases}, \quad n \in \mathbb{N}^+ \tag{65}$$

$$a_n^- = \begin{cases} P_{l-x} \left(R^+ P_l R^- P_l\right)^{\frac{n-1}{2}} R^+ P_{l-\xi}, & n \text{ is odd} \\ P_{l-x} \left(R^+ P_l R^- P_l\right)^{\frac{n}{2}} P_{\xi-l}, & n \text{ is even} \end{cases}, \quad n \in \mathbb{N} \tag{66}$$

Summing up all of the terms in Eq. (65) and Eq. (66) yields the superimposed amplitude, which by multiplying $T^{\text{inf}}(k,t)$ and applying the inverse Fourier transform yields the superimposed GF in the spatial domain:

$$T(x,\xi,t) = \left\{ \mathcal{F}^{-1}_{k \to x^{\inf}} \left[ \left( P_{-x} + P_x R^- \right) P_l \sum_{n=0}^{\infty} \left( R^+ P_l R^- P_l \right)^n \left( R^+ P_{l-\xi} + P_{\xi-l} \right) \bar{T}^{\inf}(k,t) \right] \right\}_{x^{\inf}=0} \quad (67)$$

Applying the geometric summation formula and the residue theorem to Eq. (67) yields:

$$T(x,\xi,t) = \sum_{n=-\infty}^{\infty} \frac{\left( P_{-x} + P_x R^- \right)}{4l R^+ R^- P_l} \left( R^+ P_{l-\xi} + P_{\xi-l} \right) e^{-ck_n^2 t} \quad (68)$$

where $k_n$ are roots of the characteristic function $R^+ R^- P_{2l} = 1$.

For a specific finite-domain case, suppose $T=0$ at both the left boundary $x=0$ and the right boundary $x=l$, then $R^+ = R^- = -1$, and Eq. (67) becomes:

$$\begin{aligned} T(x,\xi,t) &= \sum_{n=-\infty}^{\infty} \left[ T^{\inf}(\xi - x + 2nl, t) - T^{\inf}(\xi + x + 2nl, t) \right] \\ &= \frac{1}{\sqrt{4\pi ct}} \sum_{n=-\infty}^{\infty} \left[ \exp\left( -(\xi - x + 2nl)^2 / (4ct) \right) - \exp\left( -(\xi + x + 2nl)^2 / (4ct) \right) \right] \end{aligned} \quad (69)$$

and Eq. (68) becomes:

$$T(x,\xi,t) = \sum_{n=1}^{\infty} \frac{2}{l} \sin(k_n x) \sin(k_n \xi) e^{-ck_n^2 t} \quad (70)$$

where $k_n = n\pi / l$. Eq. (70) is the same as that obtained by the method of separation of variables.

The expression in Eq. (70) is similar to the vibration modes of a beam. However, the name 'orthogonal function' will be used to avoid ambiguity. For a semi-infinite-domain problem, if the domain lies on $x > 0$, the superimposed solution is simply:

$$T(x,\xi,t) = T^{\inf}(\xi - x, t) + R^- T^{\inf}(\xi + x, t) \quad (71)$$

The analysis in this section seems like a degenerated version of that of the Euler-Bernoulli beam. Again, reflection analysis in the Fourier transform domain perfectly links the infinite-domain, semi-infinite-domain, and finite-domain solutions. The new method also provides a new perspective that heat also conducts as propagating waves.

Another easy way to prove the equivalence between Eq. (69) and Eq. (70) is through the Poisson summation formula. It is like the equivalence between Eq. (67) and Eq. (68) reveals the essence of the Poisson summation formula. The author wonder if the equivalence between Eq. (25) and Eq. (42) is a 2-D version of the Poisson summation formula.

## 5.3 Comparison between the results calculated by two different expansion methods

$t^*$ and $\tau_l$ are defined in Eq. (31). Still assume $\xi = 0.5l$ and $x = 0.7l$. Nondimensionalize $T$ as $T^* = T \cdot l$. The two types of solutions, specifically, Eq. (69) for self-similar expansion and Eq. (70) for traditional function expansion, are compared.

The first terms and the sums of the first four terms of each type of solution are drawn in Figure 9:

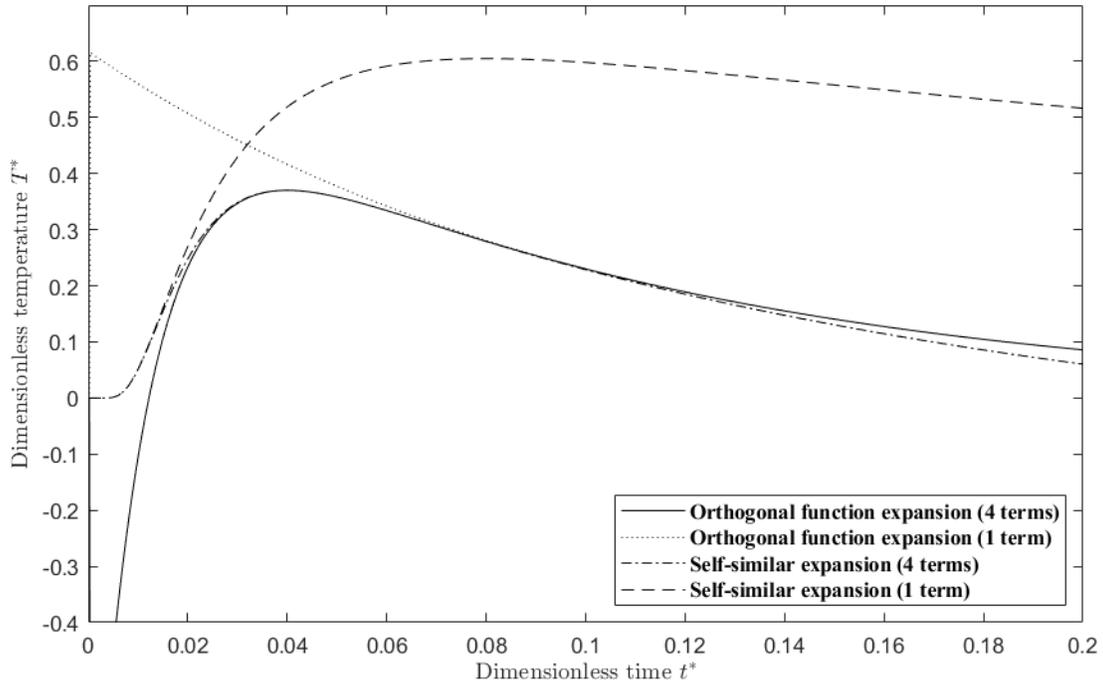

Figure 9. Comparison of two types of solutions for a one-dimensional dissipation equation problem

Like the beam case, the self-similar-expansion solution diverges at long times; however, the orthogonal function expansion does not uniformly converge but converges better at long times. The infinite-domain solution (the first term of self-similar expansion), which perfectly satisfies the governing equation under an impulse load, does not consider boundary effects; on the other hand, the first term of orthogonal expansion solution, which satisfies boundary conditions already, is a poor approximation of the impulse load.

To reach a full-time convergence, two types of solutions should be considered comprehensively,

and to obtain accurate enough solutions, each type of solution should have multiple terms. When $n=4$, the threshold time to differentiate between short time and long time is about $t^*=0.06$. For large $n$, the threshold time can be calculated analytically. Taking Eq. (69) and Eq. (70) for example, the leading terms of the decaying terms are $\exp(-n^2 l^2/ct)$ and $\exp(-c(4n)^2\pi^2 t/l^2)$, respectively. They are both proportional to $n^2$, thus converging at the same rate. Equalizing them yields the threshold of the dimensionless time:

$$t^* = 1/4\pi = 0.0795775 \tag{72}$$

## 6  Conclusions

Steady-state vibration solutions are often sought when studying dynamic responses of structures triggered by impulses. However, when truncating the solutions by ignoring high-frequency information, the solutions obtained by modal expansion converge poorly at short times. On the other hand, although the convergence of infinite-domain and semi-infinite-domain solutions are relatively accurate at initial response times, they gradually become inaccurate at relatively long times. This paper attempts to fill the gap using the self-similar wave-reflection approach in the complex Fourier transform domain to deduce a general solution applicable to various boundary conditions from the infinite-beam response and prove the wave-mode duality analytically for finite-beam Green's function. The evanescent waves are proved to possess propagating speed, and the dissipation phenomenon is generalized as wave propagation with imaginary angular frequencies.

Reflection and transmission analysis in the Fourier transform domain is proved valid in this paper. Unlike the modal expansion solutions, which converge uniformly about time, self-similar expansion solutions converge better at short response times. The newly derived modal expansion solution obtained by applying the geometric summation formula is equivalent to the traditional one, yet it avoids the calculation of the inner product of each mode, therefore more convenient and efficient. By considering both the self-similar expansion solution and modal expansion solution, the convergence of the analytical solutions is fully guaranteed. It is also found that the reciprocal of characteristic functions gives rise to imaginary impulse functions, whose magnitudes linearly determine the amplitude of each vibration mode. The semi-infinite case and the coupled beam case are also covered in this paper. The newly developed method is proved applicable to a batch of linear dispersive PDE, including the beam equation and the dissipation equation.

**Acknowledgments**

Mr. Minjiang Zhu is thankful for Southern University of Science and Technology (SUSTech), which financially supported him in studying and researching at North Carolina State University (NCSU) in Spring 2020. Mr. Zhu is also grateful for the instructions from Prof. Fuh-Gwo Yuan at NCSU and those from Prof. Peng Yu and Prof. Wei Hong at SUSTech.

**References**

[1] Graff, Karl F. Wave Motion in Elastic Solids. Columbus: Ohio State University Press, 1975. Print.

[2] B.R. Mace, "Wave reflection and transmission in beams," Journal of Sound and Vibration, Vol. 97, No. 2, pp. 237-246, 1984. https://doi.org/10.1016/0022-460X(84)90320-1

[3] C. Mei, and B.R. Mace, "Wave Reflection and Transmission in Timoshenko Beams and Wave Analysis of Timoshenko Beam Structures." Journal of Vibration and Acoustics 127.4 (2005): 382-94. Web. https://doi.org/10.1115/1.1924647

[4] F. Fahy, and P. Gardonio. Sound and Structural Vibration. 2nd ed. Burlington, Mass: Elsevier, 2007. Web. https://doi.org/10.1016/B978-0-12-373633-8.X5000-5

[5] X.Q. Wang, R.M.C. So, K.T. Chan, "Resonant Beam Vibration: A Wave Evolution Analysis." Journal of Sound and Vibration 291.3 (2006): 681-705. Web. https://doi.org/10.1016/j.jsv.2005.06.030

[6] Audoly, Basile, and NEUKIRCH, Sébastien. "Fragmentation of Rods by Cascading Cracks: Why Spaghetti Does Not Break in Half." Physical Review Letters 95.9 (2005): 095505.1-95505.4. Web. https://doi.org/10.1103/PhysRevLett.95.095505

[7] T. Akkaya, W.T. van Horssen, On constructing a Green's function for a semi-infinite beam

with boundary damping. Meccanica 52, 2251–2263 (2017). https://doi.org/10.1007/s11012-016-0594-9

[8] X. Y. Su and Y. H. Pao, "Ray-normal-mode and hybrid analysis of transient waves in a finite beam," Journal of Sound and Vibration, Vol. 152, No. 2, pp. 351-368, 1992.

[9] G.G.G. Lueschen, L.A. Bergman, D.M. McFarland, "GREEN'S FUNCTIONS FOR UNIFORM TIMOSHENKO BEAMS." Journal of Sound and Vibration 194.1 (1996): 93-102. Web. https://doi.org/10.1006/jsvi.1996.0346

[10] Ana P. Fernandes, Priscila F.B. Sousa, Valerio L. Borges, Gilmar Guimaraes, "Use of 3D-transient Analytical Solution Based on Green's Function to Reduce Computational Time in Inverse Heat Conduction Problems." Applied Mathematical Modelling 34.12 (2010): 4040-049. Web. https://doi.org/10.1016/j.apm.2010.04.006

[11] R. Buessow, "Applications of the Flexural Impulse Response Functions in the Time Domain." arXiv: Classical Physics (2006): n. pag.

[12] Sandilo SH, van Horssen WT (2012) On boundary damping for an axially moving tensioned beam. J Vib Acoust 134(1): 011005

[13] P. Folkow, P. Olsson, G. Kristensson. (1998). Time domain Green functions for the homogeneous Timoshenko beam. Quarterly Journal of Mechanics and Applied Mathematics. 51. 10.1093/qjmam/51.1.125......